\documentclass[oneside,reqno,12pt,a4paper]{amsart}
\usepackage[T1]{fontenc}
\usepackage[margin=2.5cm]{geometry}
\usepackage{bm}
\usepackage{amsmath,amsfonts,amssymb,amscd,amsthm}
\usepackage[usenames,dvipsnames,svgnames]{xcolor}
\usepackage{graphicx}
\graphicspath{}
\usepackage{caption}
\usepackage{subcaption}
\usepackage{doi}
\usepackage{makecell}
\usepackage{hyperref}
\usepackage{acronym}
\usepackage{listings}
\usepackage{textgreek}
\usepackage{url}
\usepackage{color}
\usepackage{framed}
\usepackage{verbatim}
\usepackage{fancyvrb}
\usepackage{microtype} 
\usepackage[final]{pdfpages}
\usepackage{tikz}
\usepackage{booktabs}
\usepackage{multirow}
\usepackage{algorithm}
\usepackage{algpseudocode}

\usepackage[normalem]{ulem}

\usepackage[numbers,sort&compress]{natbib}

\usepackage{textcomp}
\usepackage{cancel}
\usepackage{mathrsfs}
\usepackage{stmaryrd}
\usepackage{parskip} 
\usepackage{import}
\usepackage{xifthen}
\usepackage{pdfpages}
\usepackage{hyperref}
\usepackage{float}
\usepackage{graphicx}
\usepackage{siunitx}
\usepackage{placeins}
\usepackage[labelfont=bf,skip=10pt]{caption}
\usepackage{enumitem}
\usepackage{cleveref}
\usepackage{import}
\usepackage{transparent}
\usepackage{xcolor}
\usepackage{longtable}
\catcode`\|=12\relax
\usepackage{pgfplots}
\pgfplotsset{compat=newest}
\usepackage{program}
\usepackage{tikzscale}
\usepackage{soul}
\usepackage{pmboxdraw}

\tikzset{
every picture/.style={
line width = 0.3mm} 
}
\pgfplotsset{
every axis/.append style={
line width = 0.3mm,
grid style={
    line width = 0.3mm,
},
tick style={
    line width = 0.3mm,
},
},
}

\hypersetup{
breaklinks=true,
bookmarksopen=true,
pdftitle={Template article},    
pdfauthor={BadiaLab},     
colorlinks=true,       
linkcolor=black,          
citecolor=blue,        
filecolor=black,      
urlcolor=blue           
}
\definecolor{bg}{rgb}{0.93,0.93,0.93}

\acrodef{ode}[ODE]{ordinary differential equation}
\acrodef{pde}[PDE]{partial differential equation}
\acrodef{fe}[FE]{finite element}
\acrodef{fem}[FEM]{finite element method}
\acrodef{fcm}[FCM]{finite cell method}
\acrodef{DOF}[DOF]{degree of freedom}
\acrodefplural{DOF}[DOFs]{degrees of freedom}
\acrodef{agfem}[AgFEM]{aggregated finite element method}
\acrodef{cutfem}[CutFEM]{cut finite element method}
\acrodef{dg}[DG]{discontinuous Galerkin}
\acrodef{ls}[LS]{level set}
\acrodef{to}[TO]{topology optimization}
\acrodef{nn}[NN]{neural network}
\acrodef{ldf}[LDF]{linear driving force}
\acrodef{dac}[DAC]{direct air capture}
\acrodef{mof}[MOF]{metal organic framework}
\acrodef{tvsa}[TVSA]{temperature vacuum swing adsorption}
\acrodef{SIMP}[SIMP]{solid isotropic microstructure with penalization for intermediate densities} 
\acrodef{ipcc}[IPCC]{Intergovernmental Panel on Climate Change}
\acrodef{sipg}[SIPG]{symmetric interior penalty Galerkin}
\acrodef{supg}[SUPG]{streamline upwind Petrov-Galerkin}

\newcommand{\tnor}[1]{{\left\vert\kern-0.25ex\left\vert\kern-0.25ex\left\vert #1 
\right\vert\kern-0.25ex\right\vert\kern-0.25ex\right\vert}}

\begin{document}




\newcommand{\executeiffilenewer}[3]{%
\ifnum\pdfstrcmp{\pdffilemoddate{#1}}%
{\pdffilemoddate{#2}}>0%
{\immediate\write18{#3}}\fi%
}
\newcommand{\incfig}[2][1]{%
\executeiffilenewer{#2.svg}{#2.pdf}%
{inkscape -z -D --file=#2.svg %
--export-pdf=#2.pdf --export-latex}%
\def\svgwidth{#1\linewidth}
\input{#2.pdf_tex}%
}

\newcommand{\mycomment}[1]{}

\title[Topology optimisation of adsorption beds for cost-effective direct air capture]{Topology optimisation of adsorption beds for cost-effective direct air capture}


\author[]{Connor N. Mallon$^{1*}$}
\author[]{Aaron W. Thornton$^{2}$}
\author[]{Matthew R. Hill$^{1,2}$}
\author[]{Nora Gr\"utering$^{2}$}
\author[]{Santiago  Badia$^{3}$}
\thanks{\null\ 
$^{1}$ Department of Chemical and Biological Engineering, Monash University, Wellington Rd Clayton, 3800, Victoria, Australia.\ 
$^{2}$ CSIRO, Research Way Clayton, 3168, Victoria, Australia,\ 
$^{3}$ School of Mathematics, Monash University, Wellington Rd Clayton, 3800, Victoria, Australia.\ 
$^*$ Corresponding authors\ 
E-mails: {\tt connor.mallon@monash.edu} (Connor Mallon, Department of Chemical and Biological Engineering, Monash University, Wellington Rd Clayton, 3800, Victoria, Australia), {\tt santiago.badia@monash.edu} (Santiago Badia, School of Mathematics, Monash University, Wellington Rd Clayton, 3800, Victoria, Australia)}


\begin{abstract}
  The performance of adsorption systems is sensitive to the geometry of the adsorbent. To achieve feasibility for large-scale implementation of CO\textsubscript{2} direct air capture systems in line with the latest Intergovernmental Panel on Climate Change objectives,
significant improvements to current adsorption systems must be made. In this study, we employ a topology optimization to the macrostructure of a swing adsorption bed using an experimentally verified numerical model. We unveil a simple blunted cone design, which achieves an estimated cost range of \$49-116/t-CO\textsubscript{2} removed from the atmosphere for the system. This represents an improvement over the benchmark monolith design, 
  which incurs a cost of \$75-140/t-CO\textsubscript{2} considering the same process parameters. 
\end{abstract}


 \maketitle

\section{Introduction}

Anthropogenic climate change is one of the greatest known threats to the existence of life on Earth. The industrial emission of CO\textsubscript{2} in particular is leading to dangerous concentrations of greenhouse gases in the atmosphere. One measure to address this issue while continuing to meet society's energy demands is the development of adsorption technologies that capture and control these gases. 


One option is the scrubbing of flue gas at industrial sites. Although an important technique for minimizing industrial pollution, this technology is only suitable for large power plants which only contribute to a third of global emissions \cite{Jones2011}. An alternative that decouples the capture technology from the emission site allowing for larger scale deployment is \ac{dac} systems. The \ac{ipcc} has now highlighted the important role this technology should play in the strategies towards net-zero by 2050 \cite{IPCC2023}. Although this technology exists in various stages of development, it is mainly the cost per captured tonne of CO\textsubscript{2} that hinders large scale implementation \cite{IPCC2023}. The current cost per tonne of a DAC system is in the range of \$84–386/t-CO\textsubscript{2} \cite{NASEM2019} which must be reduced if the technology is to bridge the gap between its current state of removing CO\textsubscript{2} on the order of kt/year and its target Gt/year \cite{IPCC2023}. 

With a few exceptions, all \ac{dac} processes involve cycling reversible sorbents to capture and release CO\textsubscript{2} \cite{SanzPrez2016}. These processes are most generally categorized by the type of sorbent used and the regeneration method. Among the many alternatives, the combination selected in this work is the use of a solid sorbent material in a \ac{tvsa} process. Solid sorbents offer a high productivity and lower regeneration energy requirement compared to liquid sorbents, which can translate to a lower overall operational cost \cite{Sabatino2021}. For regeneration, the \ac{tvsa} method is technically mature, offering energy efficiency and limited drawbacks \cite{Sodiq2023,Sabatino2021}. The reader is referred to \cite{Sodiq2023,Sabatino2021,McQueen2021} for a detailed discussion of the alternative methods for \ac{dac} systems. 

Although solid sorbent \ac{tvsa} processes are considered economical compared to the alternatives, significant cost reductions are still required to reach the \ac{ipcc} targets. One way to reduce the cost is by optimizing their various components. Even after addressing the sorbent and regeneration method, many system design decisions remain without intuitive solutions. To find low-cost solutions, attempts on optimizing process parameters for solid sorbent \ac{tvsa} systems have been made \cite{Sinha2017,Luukkonen2023,Schellevis2021,Schellevis2022,Balasubramaniam2024}, which generally focus on a few scalar parameters such as inlet velocity, adsorption/desorption time and desorption temperature and pressure. 

A critical factor influencing the performance of adsorption systems is the contactor geometry. The arrangement of the sorbent within the bed significantly affects both the pressure drop and the adsorption dynamics \cite{Sandu2021}. To date, investigations into the geometry of swing adsorption systems have focused on trialing a limited number of geometrical configurations \cite{Tegeler2023,Rezaei2009} or optimizing the length and width of the encasing column only \cite{Lian2019}.

However, recent advances in additive manufacturing have significantly enhanced control over the geometry of the structured adsorbent itself \cite{Soliman2020,Lawson2021}. This increased flexibility paves the way for the discovery of novel and highly efficient contactors. To evaluate the performance of geometries that vary in multiple dimensions, we require a simulation that resolves the adsorptive material portion of the domain from the free flow domain.

When modelling adsorption systems, the relationship between fluid flow and species concentration should be considered. In most cases, one resorts to a 1D plug flow model assuming a homogeneous distribution of porous material \cite{Shafeeyan2014}. Importantly, these models cannot take into account the effect of local geometry and are therefore unsuitable when considering geometrical design of the adsorbent. Most works in multiple dimensions also only consider a single porous domain which is not resolved from a free flow region, see e.g. \cite{Lian2019,BenMansour2018,Gautier2018}, and are therefore also insufficient in cases where the entire modelled domain is not filled with the adsorbent. To capture local effects from the distribution of the material, we develop equations for both a free flow region and a porous region, allowing us to evaluate the performance of different layouts of the adsorptive material within the bed. To efficiently search through the extensive design space, we then utilize a \ac{to} strategy.


Topology optimization was born in the field of structural design \cite{Bendse1989} but has since gained popularity across various engineering disciplines, including fluid flow and species transport. Its application to adsorption systems specifically, however, is quite scarce. 
In \cite{Amigo2018-2} and \cite{Prado2021}, optimizations considering adsorption in natural gas storage systems are conducted. In these works, the entire column is filled with porous material divided into adsorptive and non-adsorptive regions. The regions of non-adsorptive material are included for thermal management in this highly exothermic application. In contrast to this application, we consider a void region where there is no material and require a fluid dynamics model to resolve the flow and transport of unadsorbed species. Thermal management in our work is not considered to be critical considering the low concentrations of CO\textsubscript{2} in ambient air and high fluid velocities through the channel, which quickly advect the exothermic heat out of the column. 

Other related works involve the topology optimization of systems involving chemical reactions in porous material. \acp{to} using steady advection-diffusion-reaction equations to solve problems in this domain can be seen in \cite{Okkels2007,Schpper2010,Yaji2017}. Outside adsorption and catalysis, other multi-physics topology optimization problems considering porous media flow have also been tackled. In \cite{Takezawa2019}, to minimize pressure drop and temperature, a fluid model for porous media is coupled to a convection-diffusion equation for heat transport. In \cite{Padhy2023}, a multi-scale topology optimization of porous media is used to maximize contact area and minimize power dissipation for fluid problems. Other works considering topology optimization for porous media problems are discussed in \cite{Alexandersen2020}. In our work, we consider the topology optimization of the macrostructure of a porous bed. We assume a microstructure and find a layout which minimizes the cost per tonne of CO\textsubscript{2} removed by the system. We consider a convective Brinkman-Forcheimer model for the fluid and a transient advection-diffusion model with adsorption for the species transport. 

This article is organized as follows. In Section \ref{problem-formulation}, we develop the physical model of the system. Then, in Section \ref{numerical-method}, we detail the numerical method used to solve this model. In Section \ref{experimental-validation}, we then validate the numerical model with a physical experiment. Next, in Section \ref{performance-evaluation}, we introduce the techno-economic model used to assess the system's performance and in Section \ref{optimization-problem} we develop the topology optimization framework based on the techno-economic model. Finally, in Section \ref{optimized-designs}, we present the optimized design and make a comparison to some benchmark solutions considering the full cost of the system. We demonstrate that by simply altering the layout of the adsorbent within the column while keeping all other process parameters fixed, we can achieve significant performance improvements compared to standard geometries.

\hypertarget{problem-formulation}{%
\section{Physical Problem formulation}\label{problem-formulation}}

A pore scale model is developed to simulate structured
adsorption systems for capturing trace CO\textsubscript{2} in ambient conditions. Equations are
formulated to solve for a velocity field \(\boldsymbol{u}\), a mechanical
pressure field \(p\), a concentration field \(c\) and an adsorbed concentration field $q$. The continuous model is detailed in this section. 



\subsection{Problem Geometry}

\begin{figure}
	\centering
  \includegraphics[width=1\linewidth]{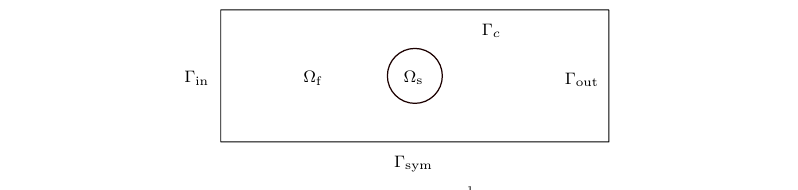}
	\caption{A schematic of the various domains within the channel.}
	\label{fig:stokes-full-channel}
\end{figure}

The domain $\Omega$, depicted in Figure \ref{fig:stokes-full-channel}, is decomposed into free flow and porous regions. In the free flow region $\Omega_\text{f}$, the flow is unobstructed. In the porous region $\Omega_{\mathrm{s}}$, there is a material scaffolding made up of a particular adsorbent. The channel inlet is denoted with $\Gamma_\text{in}$, the outlet with $\Gamma_\text{out}$ and the remaining surface with $\Gamma_c$. We design the geometry such that $ (\Gamma_\text{in} \cup \Gamma_\text{out}) \cap \Omega_{\mathrm{s}} = \emptyset$. The problem is considered axisymmetric and we impose appropriate symmetric conditions on $\Gamma_\text{sym}$.

\hypertarget{flow-problem}{%
\subsection{Flow Problem}\label{flow-problem}}

In the free flow region, the steady incompressible Navier-Stokes equations are used:
\begin{equation}\protect\hypertarget{eq:u}{}{
  \left\lbrace   
  \begin{aligned}
   \rho (\boldsymbol{u}\cdot (\boldsymbol{\nabla}\boldsymbol{u}) )
   - \boldsymbol{\nabla}\cdot (\mu \boldsymbol{\nabla} \boldsymbol{u}) +\boldsymbol{\nabla}p&= \boldsymbol{0} &&\text{in } \Omega_\text{f}, \\
  \boldsymbol{\nabla} \cdot \boldsymbol{u}&= 0 &&\text{in } \Omega_\text{f}, \\
  \end{aligned}
  \right.   
}\label{eq:u}\end{equation}
where $\mu$ is the fluid viscosity and $\rho$ is the fluid density. To model the flow in the porous region, the convective Brinkman-Forcheimer equation is used:
\begin{equation}\protect\hypertarget{eq:u}{}{
  \left\lbrace   
  \begin{aligned}
    \rho (\boldsymbol{u}\cdot (\boldsymbol{\nabla}\boldsymbol{u}) )
   - \boldsymbol{\nabla}\cdot (\mu \boldsymbol{\nabla} \boldsymbol{u}) +
   \alpha \boldsymbol{u} + \beta \vert \boldsymbol{u} \vert \boldsymbol{u}
  +\boldsymbol{\nabla}p
  &= \boldsymbol{0} &&\text{in } \Omega_\text{s}, \\
  \boldsymbol{\nabla} \cdot \boldsymbol{u}&
  = 0 &&\text{in } \Omega_\text{s}, \\
  \end{aligned}
  \right.   
  }\label{eq:u}\end{equation}
where $\alpha$ is the Darcy coefficient and $\beta$ is the Forchemier coefficient. {We consider transmission conditions at the fluid-solid interface:
\begin{equation}
  \begin{aligned}
\boldsymbol{u}_{-} &= \boldsymbol{u}_+ \quad &\text{on}\  \partial\Omega_{\mathrm{s}}\cap\partial\Omega_{\mathrm{f}}, \\
\mu \nabla \boldsymbol{u}_- \cdot \boldsymbol{n} &= \mu \nabla \boldsymbol{u}_+ \cdot \boldsymbol{n} \quad &\text{on}\  \partial\Omega_{\mathrm{s}}\cap\partial\Omega_{\mathrm{f}},
  \end{aligned}
\end{equation}
where $\boldsymbol{u}_-$ and $\boldsymbol{u}_+$ are the values of $\boldsymbol{u}$ on either side of the interface and $\boldsymbol{n}$ is the outward normal on $\partial \Omega_{\mathrm{s}}$.}
A parabolic velocity profile $\boldsymbol{u}_\text{in}$ is imposed on the inlet:
\begin{equation}\label{eq:uin}
  \boldsymbol{u} = \boldsymbol{u}_\text{in} \quad \text{on}\  \Gamma_{\text{in}},
\end{equation}
a zero flux condition at the outlet
\(\Gamma_{\mathrm{out}}\): \begin{equation}\protect\hypertarget{eq:u_out}{}{
\mu \boldsymbol{\nabla} \boldsymbol{u} \cdot \boldsymbol{n} - p\boldsymbol{n} = \boldsymbol{0} \quad \text{on}\ \Gamma_{\text{out}},
}\label{eq:u_out}\end{equation} 
and a no-slip condition is prescribed on
the channel boundary in the fluid region: \begin{equation}
\boldsymbol{u}=\boldsymbol{0} \quad\text{on}\ \Gamma_{\text{c}}\cap\partial\Omega_{\mathrm{f}}.
\label{eq:u_s}\end{equation}
Note that we do not impose the no slip condition for the porous region.
Finally, a symmetry condition is imposed on the symmetry boundary $\Gamma_\text{sym}$:
\begin{equation}\label{eq:usym}
\boldsymbol{u}\cdot \boldsymbol{n} = \boldsymbol{0} \quad \text{on}\ \Gamma_{\text{sym}}.
\end{equation}


\hypertarget{species-transport}{%
\subsection{Species Transport}\label{species-transport}}

To model the species concentration field in the free flow domain
\(\Omega_\text{f}\), the transient advection-diffusion equation is used:
\begin{equation}\protect\hypertarget{eq:c}{}{
\frac{\partial c}{\partial t} + \boldsymbol{u} \cdot \boldsymbol{\nabla}c
- \boldsymbol{\nabla} \cdot ( D_\text{f} \boldsymbol{\nabla}c )= 0\quad\text{in}\ \Omega_\text{f},
}\label{eq:c}\end{equation} where \(c\) is the concentration in the
volume and \(D_\text{f}\) is the diffusion coefficient for the fluid. 
In the porous domain, we also consider a mass sink term from adsorption:
\begin{equation}\protect\hypertarget{eq:c}{}{
  \epsilon \frac{\partial c}{\partial t} + \boldsymbol{u} \cdot \boldsymbol{\nabla}c
  - \boldsymbol{\nabla} \cdot ( D_\text{s} \boldsymbol{\nabla}c ) + 
  (1-\epsilon) \frac{\partial q}{\partial t}
  = 0 \quad\text{in}\ \Omega_\text{s},
  }\label{eq:cs}\end{equation} where $D_\text{s}$ is the effective diffusion coefficient for the porous region, $\epsilon$ is the porosity and $q$ is the adsorbed concentration. The linear driving force model is used to model the adsorption rate:
  \begin{equation}
    \frac{\partial q}{\partial t} = k (q^*(c)-q)\quad\text{in}\ \Omega_\text{s},
  \end{equation}

  where $k$ is the mass transfer resistance and $q^*(c)$ is the linear isotherm $q^*=Kc$. Here, $K$ is determined experimentally.
  Initially, both the unadsorbed and adsorbed concentration in the fluid and solid domains are zero:
\begin{equation}
  c\vert_{t=0} = m\vert_{t=0} = 0 \quad \text{in} \  \Omega_{\text{f}}\cup\Omega_{\text{s}}.
\end{equation}
For the boundary conditions, a constant concentration profile equal to the ambient CO\textsubscript{2} concentration
is imposed at the inlet boundary \(\Gamma_{in}\):
\begin{equation}\protect\hypertarget{eq:c_in}{}{
c= \bar{c}\quad\text{on}\ \Gamma_{in},
}\label{eq:c_in}\end{equation} and a zero flux condition is used at the
outlet \(\Gamma_{out}\): \begin{equation}\protect\hypertarget{eq:c_out}{}{
\boldsymbol{\nabla} c\cdot \boldsymbol{n} = 0 \quad \text{on}\ \Gamma_{out}.
}\label{eq:c_out}\end{equation} 

A zero flux condition is also imposed on the channel surface $\Gamma_\text{c}$:
\begin{equation}
  \boldsymbol{\nabla} c\cdot \boldsymbol{n} = 0 \quad \text{on}\ \Gamma_{\text{c}},
\end{equation}

and the symmetry boundary $\Gamma_\text{sym}$:
\begin{equation}
  \boldsymbol{\nabla} c\cdot \boldsymbol{n} = 0 \quad \text{on}\ \Gamma_{\text{sym}}.
\end{equation}

\subsection{Pellet Parameters}

The solid zones in the optimized substrate designs have the geometry of packed pellets. Due to this, we can leverage some well established empirical models designed for packed beds. We consider the standard approximation of the Darcy and Forchemier coefficients for packed beds \cite{Ergun1952FluidFT}, which are a function of the porosity $\epsilon$ and pellet diameter $d_{p}$:
\begin{equation}
\alpha = \frac{150(1-\epsilon)^2}{d_{p}^2 \epsilon^3} \mu 
\label{eq:darcy-coefficient}
\end{equation}
\begin{equation}
  \beta = \frac{1.75(1-\epsilon)}{d_{p}\epsilon^3}\rho
  \label{eq:forcheimer-coefficient}
\end{equation}

This model is used for both cylindrical and spherical pellets. For the adsorption rate, we consider the dominant mass transfer resistance, namely, the intraparticle resistance. For spherical pellets, this is given by \cite{Glueckauf1955}:
\begin{equation}
  k = \frac{30 D_s}{ d_p},
\end{equation}
and for cylindrical pellets, by \cite{Patton2004}: 
\begin{equation}
k = \frac{16 D_s}{ d_p}.
\end{equation}

\subsection{Monolith Parameters}\label{monolith-parameters}

When modelling packed pellets, the porosity used in \eqref{eq:darcy-coefficient} and \eqref{eq:forcheimer-coefficient} comes from the voids between the pellets. When modelling the monolithic structure, the solid adsorbent region is considered to be completely full of adsorbent material. The macro-porosity in this case is therefore zero. What this means for the flow is that fluid cannot penetrate into the solid domain. To deal with this situation, we set $\alpha$ high enough so that the flow is completely stopped in $\Omega_{\text{s}}$. For the transport, the model becomes a pore-diffusion model, since we now instead explicitly model the intraparticle diffusion in space using $D_s$ and consider the micro-porosity of the sorbent. In the pore diffusion model, since we are explicitly modelling the intraparticle resistance, we can assume that there is no further resistance to a particle being adsorbed, i.e. $k\rightarrow \infty$ and therefore $q=q^*$ everywhere in $\Omega_{\text{s}}$. We can therefore eliminate $q$ and take the adsorption rate as:
\begin{equation} 
   \frac{\partial{q}}{\partial{t}} = \frac{\partial{q^*}}{\partial{c}} \frac{\partial{c}}{\partial{t}}.
\end{equation}

\newcommand{\vbs}{\boldsymbol{v}}
\newcommand{\nbs}{\mathbf{n}}

\hypertarget{numerical-method}{%
\section{Numerical Method}\label{numerical-method}}

\hypertarget{discrete-spaces}{%
\subsection{Discrete Spaces}\label{discrete-spaces}}

The \ac{fe} method used to solve the problem is described in this section. Let $\mathcal{T}_h$ represent a cartesian mesh of $\Omega \doteq \Omega_{\text{s}} \cup \Omega_{\text{f}}$ with a characteristic mesh size $h$. We define a nodal Lagrangian \ac{fe} space of order $k$ on $\mathcal{T}_h$ as:
\begin{equation}
  V_{h}^k = \{ v_h \in H^1(\Omega): v_h\vert_{K} \in Q_k \ \forall K \in \mathcal{T}_h \},
\end{equation}
where $Q_k$ is the space of polynomials with maximum degree $k$ in each variable. We also define the discontinuous space of order $k$ as:
\begin{equation}
  V_{h}^{\text{disc},k} = \{ v_h \in L^2(\Omega): v_h\vert_{K} \in Q_k \ \forall K \in \mathcal{T}_h \}, 
\end{equation}
and finally, the Raviart-Thomas space \cite{ern_theory_2004} of order $k$ as:
\begin{equation}
\boldsymbol{V}_h^{RT,k}  = \{  \boldsymbol{v}_h  \in H(\text{div},\Omega): \boldsymbol{v}_h \vert_K \in [Q_{k-1}(K)] 
^{d} \oplus \boldsymbol{x} \ Q_{k-1}(K) \ \forall K \in \mathcal{T}_h \}, 
\end{equation}

\hypertarget{weak-formulation}{%
\subsection{Weak formulation}\label{weak-formulation}}

The weak formulation of the problem solved by the method is now described. The problem is composed of forms to
be integrated on the various triangulations of the problem and are
formulated in the remainder of this section. We initially refrain from defining the specific terms and instead first introduce the total residual.
The coupling between the flow and transport is unidirectional, allowing us to first solve the fluid dynamics problem and subsequently solve the transport equations.
To resolve the flow, we find \((\boldsymbol{u}_h,p)\in \boldsymbol{V}_{h,g}^{RT,1} \times V_h^{\text{disc},1} \)
such that: \begin{equation}\protect\hypertarget{eq:rins}{}{
\mathscr{R}_{\boldsymbol{u}_h,p_h}^\Omega + \mathscr{R}_{\boldsymbol{u}_h,p_h}^\Gamma + \mathscr{R}_{\boldsymbol{u}_h,p_h}^{\Omega,\text{SIPG}} + \mathscr{R}_{\boldsymbol{u}_h,p_h}^{\Omega,\text{SUPG}} =0
}\label{eq:rins}\end{equation}
for all \((\boldsymbol{v}_h,\theta_h)\in \boldsymbol{V}_{h,0}^{RT,1} \times V_h^{\text{disc},1} \) where {$\boldsymbol{V}_{h,g}^{RT,1}$ represents the set of functions in $\boldsymbol{V}_h^{RT,1}$ which satisfy the normal component of the Dirichlet conditions \eqref{eq:uin}, \eqref{eq:u_s} and \eqref{eq:usym} and $\boldsymbol{V}_{h,0}^{RT,1}$ represents the set of functions in $\boldsymbol{V}_h^{RT,1}$ which have a zero normal component on $\Gamma_{\mathrm{in}}$, $\Gamma_{\mathrm{out}}$ and $\Gamma_{\mathrm{sym}}$}
. 

For the use of a density based \ac{to} strategy, we require a numerical method that is stable in both the Darcy and Navier-Stokes regimes. We refer to \cite{Mardal2002} for a non-conforming \ac{fe} pair stable in both regimes and to \cite{Badia2009} for a stabilized \ac{fe} formulation. In this, work, we consider a div-conforming \ac{fe} space suitable for the Darcy regime combined with a discontinuous Galerkin method to make it suitable for the Navier-Stokes regime \cite{KNN2011-2}. In particular, the mixed space $\boldsymbol{V}_{h}^{RT,1} \times V_h^{\text{disc},1}$ is inherently stable in the Darcy region \cite{ern_theory_2004} and is also stable in the Navier-Stokes region with the addition of \ac{dg} terms that penalize jumps of tangent traces on element boundaries \cite{Wang2009}. In contrast to nodal langrangian \ac{fe} methods, this  formulation ensures pointwise divergence-free solutions.

In the second stage of the problem, we solve the advection-diffusion
equation and find $(c_h,q_h) \in V_{h,g}^1  \times V_{h}^{\text{disc},0}$ such that:
\begin{equation}\protect\hypertarget{eq:rad}{}{
\mathscr{R}_{c_h,q_h}^\Omega + \mathscr{R}_{c_h,q_h}^{\Omega,\text{SUPG}} = 0
}\label{eq:rad}\end{equation} 
for all
$w_h\in V_{h,0}^1 \times V_{h}^{\text{disc},0} $. 

\subsubsection{Domain Relaxation}\label{domain-relaxation}
We develop our weak form so that it can be used within a topology optimization framework. Rather than have separate integrals for the free flow and solid regions, we combine them using an indicator field to alternate between the two, allowing for the use of a density-based topology optimization strategy. A value of $\psi=1$ is used for the portion of the design domain consisting of porous material and a value of $\psi=0$ is used for the unobstructed region. 

\hypertarget{fluid-interior-terms}{%
\subsubsection{Flow Terms}\label{fluid-terms}} 
{For both the free flow and porous regions, we utilize a variational form of the convective Brinkman-Forcheimer equations. By using the indicator function as a coefficient for the Darcy and Forcheimer terms, we obtain a formulation that reduces to the correct regimes in both regions. Specifically, in the free flow region, the indicator field is zero and the Darcy and Forcheimer terms vanish, leaving only Navier-Stokes terms. In the porous region, all the terms remain and we require a solution of the full convective Brinkman-Forcheimer equations. Using this approach, the weak form of the bulk terms of the system can be expressed as:}
\begin{equation}\protect\hypertarget{eq:rins_ux3a9}{}{
\begin{aligned}
\mathscr{R}_{\boldsymbol{u}_h,p_h}^\Omega \doteq 
\int_\Omega
\ [
&\psi \left( \alpha \boldsymbol{u}_h\cdot \boldsymbol{v}_h + 
            \beta \vert \boldsymbol{u}_h \vert \boldsymbol{u}_h\cdot \boldsymbol{v}_h  \right) \\
&+ \rho \boldsymbol{v}_h  \cdot ( 
(\boldsymbol{u}_h\cdot\boldsymbol{\nabla}) \boldsymbol{u}_h ) 
+\mu( \boldsymbol{\nabla}\boldsymbol{u}_h:\boldsymbol{\nabla}\boldsymbol{v}_h)  
+p_h (\boldsymbol{\nabla} \cdot \boldsymbol{v}_h) 
+(\boldsymbol{\nabla} \cdot \boldsymbol{u}_h) \theta_h 
] 
{\rm d}\Omega.  \\
\end{aligned}
}\label{eq:rins_ux3a9}
\end{equation}
For conformity and stability of the discontinuous formulation, we utilize a \ac{sipg} formulation \cite{Knnn2011}:
\begin{equation}
  \begin{aligned}
    \mathscr{R}_{ \boldsymbol{u}_h, \boldsymbol{v}_h }^{\Omega,\text{SIPG}} \doteq
  \sum_{F\in\mathcal{F}_h} \int_{F} \ 
\bigg[ & \frac{\sigma_\mu}{h_F} \llbracket\boldsymbol{u}_h\rrbracket\cdot\llbracket\boldsymbol{v}_h\rrbracket        
    -\llbracket \boldsymbol{u}_h\rrbracket \cdot \{\!\!\{ \mu \boldsymbol{\nabla} \boldsymbol{v}_h \}\!\!\} \boldsymbol{n}
    -\{\!\!\{ \mu \boldsymbol{\nabla} \boldsymbol{u}_h \}\!\!\} \boldsymbol{n} \cdot  \llbracket\boldsymbol{v}_h \rrbracket 
\bigg]\text{d}F.
\label{eq:rins_ux3a9}
  \end{aligned}
\end{equation}
where $\sigma_\mu$ is a tunable parameter given in Table \ref{tab:stabilization-parameters}. On the boundary, we impose the remaining Dirichlet data weakly via Nitsche's method and include the right hand side term related to the \ac{sipg} formulation:
\begin{equation}\protect\hypertarget{eq:ru\Gamma}{}{
  \begin{aligned}
  \mathscr{R}_{\boldsymbol{u}_h,p_h}^\Gamma \doteq 
  \int_{\Gamma_{\text{in}}\cup (\Gamma_c \cup \Omega_\text{f})}
  \bigg[ 
  &\frac{\gamma}{h}(\boldsymbol{u}_h-\boldsymbol{g}) \cdot \boldsymbol{v}_h
  - (\boldsymbol{u}_h-\boldsymbol{g}) \cdot ( \mu \boldsymbol{\nabla}\boldsymbol{v}_h\cdot \boldsymbol{n})\\
  &- (\mu \boldsymbol{\nabla}(\boldsymbol{u}_h)\cdot \boldsymbol{n})\cdot \boldsymbol{v}_h
  + \frac{\sigma_\mu}{h} \boldsymbol{g}\cdot\boldsymbol{v}_h
  \bigg]\text{d}\Gamma,
  \end{aligned}
  }\label{eq:rins_ux3a9}\end{equation}
where $\gamma$ is a tunable parameter given in Table \ref{tab:stabilization-parameters}. We also use a standard \ac{supg} method \cite{Brooks1982} to prevent spurious oscillations in the fluid velocity due to the convective term.  
\begin{equation}
    \begin{aligned}
    \mathscr{R}_{ \boldsymbol{u}_h, \boldsymbol{v}_h }^{\Omega,\text{SUPG}} \doteq
    \int_\Omega 
    [
	   &(\tau_{\boldsymbol{u}} (\boldsymbol{u}_h \cdot \boldsymbol{\nabla}) \boldsymbol{v}_h) 
    \cdot( 
    \psi ( \alpha \boldsymbol{u}_h  + \beta \vert \boldsymbol{u}_h \vert \boldsymbol{u}_h ) 
    + 
    \rho(\boldsymbol{u}_h\cdot\boldsymbol{\nabla}) \boldsymbol{u}_h \\
    &- \boldsymbol{\nabla}\cdot(\mu\boldsymbol{\nabla}\boldsymbol{u}_h)
    + \boldsymbol{\nabla} p_h 
    )
     ] \ 
    {\rm d}\Omega  
    \end{aligned}
\end{equation}
where the SUPG parameter \(\tau_{\boldsymbol{u}}\) is taken as in \cite{Tezduyar1992}:
\begin{equation}
\begin{aligned}
  \tau_{\boldsymbol{u}} &\doteq \alpha_{\tau,\boldsymbol{u}} \left[ \left( \frac{2\vert\boldsymbol{u}_h\vert }{h} \right)^2 + 9 \left( \frac{4\mu}{\rho h^2} \right)^2 \right]^{-1/2}, \\
\end{aligned}
\end{equation}
where \(\alpha_{\tau,\boldsymbol{u}}\) is a tunable parameter given in Table \ref{tab:stabilization-parameters}.

\hypertarget{transport-terms}{%
\subsubsection{Transport Terms}\label{transport-terms}}

In the bulk, we use the indicator field to switch between the advection-diffusion with and without adsorption in the free flow and porous regions. We refrain from including the simplifications possible in the monolith case described in Section \ref{monolith-parameters} and present only the general form:
\begin{equation}
    \begin{split}
    \begin{aligned}  
      \mathscr{R}^{\Omega}_{c_h,q_h}
      \doteq
      \ \ \int_{\Omega}
      \bigg[
      &\psi
      \bigg(
        w_h \bigg( \epsilon \frac{\partial c_h}{\partial t} + (1-\epsilon) (k(q^*(c_h) - q_h)) +  \boldsymbol{u}_h \cdot \boldsymbol{\nabla}c_h  \bigg)  + D_\text{s} \boldsymbol{\nabla}c_h \cdot \boldsymbol{\nabla} w_h   \\
        &+ g_h\bigg( \frac{\partial q_h }{\partial {t}} - (k(q^*(c_h) - q_h)) \bigg)\bigg) 
      \\
       &+ (1-\psi) 
      \left(
        w \bigg( \frac{\partial c_h}{\partial t} +  \boldsymbol{u}_h \cdot \boldsymbol{\nabla}c_h  \bigg) + D_\text{f} \boldsymbol{\nabla} c_h \cdot \boldsymbol{\nabla} w_h + g_h \frac{\partial q_h }{\partial {t}} 
  \right) 
       \bigg]
       \, \mathrm{d}\Omega. \\
      \end{aligned}
    \end{split}
    \label{eq:adit}
    \end{equation}
We also include a \ac{supg} stabilization in the free flow region to stabilize against spurious oscillations in the concentration due to advection: 
\begin{equation}
    \begin{aligned}
    \mathscr{R}_{c_h,q_h}^{\Omega,\text{SUPG}}
    \doteq                                                    
    \int_{\Omega} \bigg[ \ 
    &(  \tau_c  \boldsymbol{\nabla} w_h  ) \cdot
    \bigg( \psi \left( \epsilon \frac{\partial c_h}{\partial t} + (1-\epsilon)(k(q^*(c_h) - q_h)) + \boldsymbol{u}_h \cdot \boldsymbol{\nabla} c_h - \boldsymbol{\nabla} \cdot ( D \boldsymbol{\nabla} c_h ) \right)  \\
    &+ (1-\psi) \left( \frac{\partial c_h}{\partial t} + \boldsymbol{u}_h\cdot \boldsymbol{\nabla}c_h 
    - \boldsymbol{\nabla} \cdot ( D_f \boldsymbol{\nabla} c_h )
    \right) \bigg)
    \ \bigg] {\rm d}\Omega.
  \end{aligned}
\end{equation}
The \ac{supg} parameter \(\tau_{c}\) is:
\begin{equation} 
  \tau_{c} \doteq \alpha_{\tau,c} \ \left[ \left( \frac{2\vert \boldsymbol{u}_h\vert}{h} \right)^2 + 9 \left( \frac{4D}{h^2} \right)^2 \right]^{-1/2} \\  
\end{equation}
where $\alpha_{\tau,c}$ is a tunable parameter given in Table \ref{tab:stabilization-parameters}.

\begin{table}[htbp]
  \centering
  \caption{Stabilization parameters}
  \begin{tabular}{ll}
  \hline
  Constant & Value \\
  \hline
  $\gamma$ & $10^3$ \\
  $\sigma_\mu$ & $10^5$ \\
  $\alpha_{\tau,\boldsymbol{u}}$ & 1 \\
  $\alpha_{\tau,c}$ & 1 \\

  \hline
  \end{tabular}
  \label{tab:stabilization-parameters}
\end{table}

\section{Experimental Validation}\label{experimental-validation}

To validate the numerical model, we conducted breakthrough experiments in the laboratory using both a monolithic contactor and a typical packed bed of pellets. Our proposed optimized design integrates elements of these two configurations. Therefore, verifying the breakthrough results for these setups suggests that similar accuracy can be expected for the optimized design numerical solutions. In this section, we describe the experimental methodology and then compare the experimental and numerical results.

\subsection{Experimental Method}

\subsubsection{Materials}

The zeolite 13X in powder form, which is used for the monolith coating, is supplied by Sigma Aldrich and has an average particle size of 2$\mu$m. The zeolite 13X pellets, also supplied by Sigma Aldrich, have a cylindrical shape and a diameter of 3.2mm. The cylindrical monolith is purchased in a diameter of 250mm and a length of 150mm. It is then cut to fit inside the tube, which has a diameter of 9.4mm and a length of 119mm. The tube that is filled with pellets has a diameter of 9.4mm and a length of 11.7mm.

\subsubsection{Coating Suspension}

For the monolith experiment, a coating suspension is prepared by first adding a binder to a solvent and stirring on a roller mixer for at least 24 hours. Then, zeolite 13X is added, and the mixture is stirred on the roller mixer for an additional 48 hours to ensure a homogeneous texture. The suspension contains 40 wt.\% of zeolite 13X and 5 wt.\% of binder per gram of zeolite.

\subsubsection{Coating Method}

The monolith substrate is coated using a dip-coating method. Initially, the substrate is cleansed in an ultrasonic bath with a mixture of DI water and liquid pyroneg, followed by a final rinse with DI water. It is then dried at 120°C for at least 12 hours in an oven. For the coating process, the substrate is dipped into the coating suspension and then briefly dried with compressed air to clear any blocked pores from excess suspension. After this, the substrate is placed in a vacuum oven and dried at 100°C for a minimum of 24 hours to ensure thorough drying. Throughout the process, the coating suspension is continuously stirred on a roller mixer to maintain a homogeneous mixture.

\subsubsection{Scanning Electron Microscopy}

The morphology of the coating layer on the monolith substrate is analyzed using scanning electron microscope (SEM) imaging. To achieve a clean break, the substrate is first immersed in liquid nitrogen. It is then mounted on an aluminum stub using double-sided conductive carbon tape. The substrate undergoes iridium coating using a Cressington 208HRD, with the thickness of the iridium layer being approximately 4 nm (applied at 60 mA for 50 seconds). This conductive coating is essential to prevent charge accumulation, which is crucial for obtaining clear images in an electron microscope, especially with insulating materials. The sample is imaged using a Zeiss Merlin Field Emission Scanning Electron Microscope (FESEM) operated in the secondary electron mode at an accelerating voltage of 5 kV. This gives an accurate measure for the thickness of the adsorbent layer on the monolith.

\subsubsection{Isotherms}

The samples of pellets and powder are filled into the pre-weighted sample tubes and are degassed for 24 h at 200\textdegree C. Then, the samples are weighed to determine the mass of the activated adsorbent. CO\textsubscript{2} adsorption isotherms are collected at 25\textdegree C in a range of 0.015 mbar to 1200 mbar. 

\subsubsection{Testing Rig}

\begin{figure}
	\centering
  \includegraphics[width=1\linewidth]{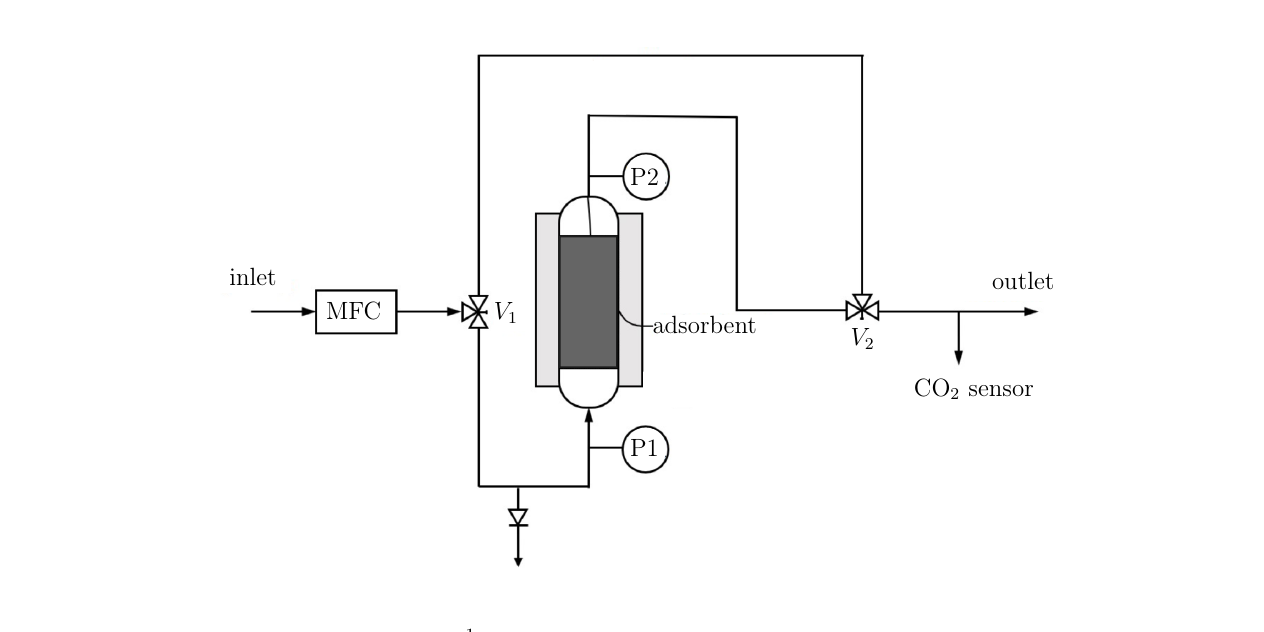}
	\caption{Testing rig schematic}
	\label{fig:experimental-setup}
\end{figure}

The samples are placed in the 9.4mm column and the column is connected to the test rig made of 0.25 inch tubes by two Swagelok fittings on both ends as seen in Figure \ref{fig:experimental-setup}. A SevenStar CS-2000 mass flow controller (MFC) meters the incoming gas and connects the gas bottles to the testing rig. The MFC is connected to a SevenStar D08-1F flow readout box that is controlled by the multi-digital MFC software v1.00 from SevenStar. Between the MFC and column, a valve (V1) is placed. The valve makes it possible to switch between the column and bypass line. Two pressure indicators are installed at the inlet and outlet of the column. The pressure indicator at the inlet of the column (P1) is an Omega PX119-1.5KAI high pressure indicator with a range from 0 to 1500 psia. The second pressure indicator(P2) is an Omega PX119-030AI low pressure indicator for pressure ranges from 0 to 30 psia. Valve V2 is placed downstream of the column, which can be switched between the bypass and column line. The outlet of the testing rig is connected to a SCD30 CO\textsubscript{2} sensor supplied by Sensirion. The CO\textsubscript{2} measurement range is 0 to 40000 ppm with an accuracy of 30 ppm.

\subsubsection{Pressure Drop Measurements}
The pressure drop is measured for the pellets by placing the sample within the 119mm long tube and switching valves V1 and V2 to the column. The experiment is repeated for a variety of inlet velocities between 0 and 1.2 ms\textsuperscript{-1}.

\subsubsection{Breakthrough Measurements}\label{btm}

A premix gas bottle of 526ppm CO\textsubscript{2} in N2 is connected to the input. The feed line is opened and the gas  mixture is flushed through the bypass until a CO\textsubscript{2} concentration is measured at the rig outlet. Then, V1 and V2 are switched to the column and the adsorption measurements are aborted after 2.5 hours.

\subsection{Validation}\label{validation}

{The Julia software library Gridap \cite{Verdugo2019a,Verdugo2022} was used to implement the \ac{fe} discretization and obtain numerical solutions to the governing equations.  It provides a high-level API for the definition of the weak form for seamless implementation of the problems defined in \eqref{eq:rins} and \eqref{eq:rad}. Furthermore, being 100$\%$ Julia code, the \ac{fe} machinery can be easily differentiated and combined with gradient-based optimization.}

\subsubsection{Isotherms}

\begin{table}[htbp]
  \centering
  \caption{Fitted isotherm parameters}
  \begin{tabular}{lll}
  \hline
  Constant & Value & Unit \\
  \hline
  Pellets Henry constant, $K_\text{pel}$ & 0.59 & mmol g\textsuperscript{-1} mbar\textsuperscript{-1} \\
  Monolith coating Henry constant, $K_\text{mon}$ & 0.62 & mmol g\textsuperscript{-1} mbar\textsuperscript{-1} \\
  \hline
  \end{tabular}
  \label{tab:geometrical-constants}
\end{table}

\begin{figure}[H]%
  \centering
  \begin{subfigure}{0.5\textwidth}
  \centering
  \includegraphics[width=1\linewidth]{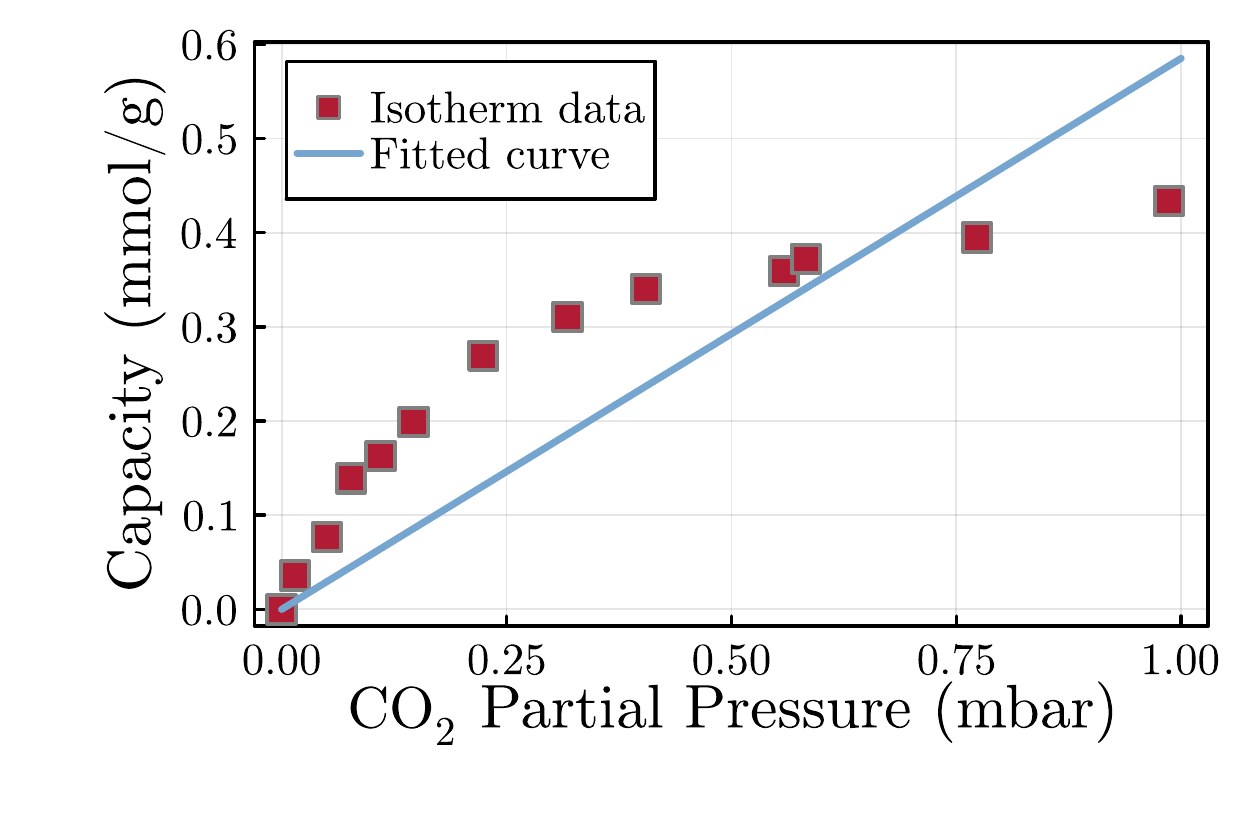}
      \caption{Pellets}
      \label{fig:os_1}
  \end{subfigure}%
  \begin{subfigure}{0.5\textwidth}
  \centering
  \includegraphics[width=1\linewidth]{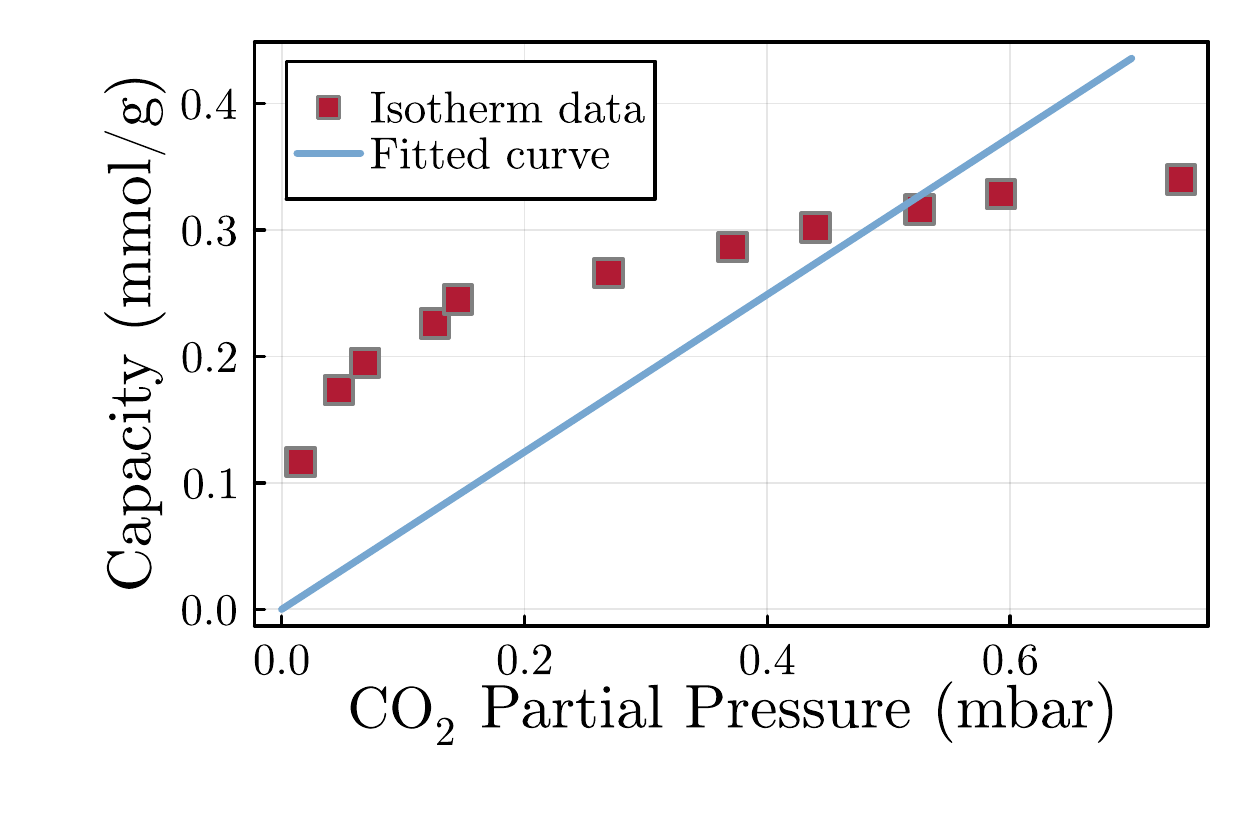}
      \caption{Monolith coating suspension}
      \label{fig:os2}
  \end{subfigure}
 
  \caption{Adsorption isotherm fitting}
  \label{fig:isotherms}
\end{figure}

For an accurate numerical model, we must capture the correct adsorption capacity for the material. To do this, we take isotherm measurements for both the pellets and the powder and binder mix used for the monolith and fit a suitable isotherm. As we only see low partial pressures of CO\textsubscript{2}, a Henry isotherm is fitted to the data in both cases. The fitted isotherms can be seen in Figure \ref{fig:isotherms}.  

\subsubsection{Breakthrough Experiments}

\begin{table}[htbp]
  \centering
  \caption{Geometrical Constants for Validation}
  \begin{tabular}{lll}
  \hline
  Constant & Value & Unit \\
  \hline
  Cylinder inner radius & 9.4 & mm \\
  Monolith length in channel & 119 & m \\
  Monolith adsorption layer thickness & 75 & $\mu$m \\ 

  Packed bed length for breakthrough & 11.7 & mm \\
  Packed bed length for pressure drop & 119 & m \\

  Pellet radius, $R_p$ & 3.2 & mm \\

  Particle porosity, $\epsilon_{\text{par}}$ & 0.37 & m \\
  Packed bed macro porosity, $\epsilon_{\text{bed}}$ & 0.31 & m \\
  
  \hline
  \end{tabular}
  \label{tab:geometrical-constants}
\end{table}

\begin{table}[htbp]
  \centering
  \caption{Physical Constants for Validation}
  \begin{tabular}{lll}
  \hline
  Constant & Value & Unit \\
  \hline
  Inlet velocity, $u_{\text{in}}$  & 60 & mm s\textsuperscript{-1} \\
  Darcy coefficient, $\alpha$  & 3651 & kg mm\textsuperscript{-3} s \textsuperscript{-1} \\
  Forcheimer coefficient, $\beta$  & 11 & kg mm\textsuperscript{-4} s \\
  Air density, $\rho$ & 1.184(10\textsuperscript{-9}) & kg mm\textsuperscript{-3} \\
  Air viscosity, $\mu$ & 1.846(10\textsuperscript{-8}) & m \\
  Inlet CO\textsubscript{2} concentration, $c_\text{in}$ & 0.0215 & nmol mm\textsuperscript{-3} \\

  Air diffusivity, $D_g$ & 16 & mm\textsuperscript{2} s\textsuperscript{-1} \\  

  Zeolite pore diffusivity, $D_m$ & 4(10\textsuperscript{-3}) & mm\textsuperscript{2} s\textsuperscript{-1} \\
  Zeolite effective diffusivity, $D_e$ & 1.16 & mm\textsuperscript{2} s\textsuperscript{-1} \\
  Mass Transfer Coefficient, k & 0.0125 & s\textsuperscript{-1} \\
  Zeolite density, $\rho_a$ & 1.06(10\textsuperscript{-6}) & kg mm\textsuperscript{-3} \\

  \hline
  \end{tabular}
  \label{tab:physical-constants}
\end{table}

\begin{table}[htbp]
  \centering
  \caption{Numerical Constants for Validation}
  \begin{tabular}{lll}
  \hline
  Constant & Value & Unit \\
  \hline
  packed bed mesh size, $h_{\text{bed}}$ & 0.094 & mm \\
  monolith mesh size, $h_{\text{mon}}$ &   7(10\textsuperscript{-3}) & mm \\
  time step, $dt$ & 5 & m \\
  \hline
  \end{tabular}
  \label{tab:numerical-constants}
\end{table}

\begin{figure}[H]%
  \centering
  \begin{subfigure}{0.5\textwidth}
  \centering
  \includegraphics[width=1\linewidth]{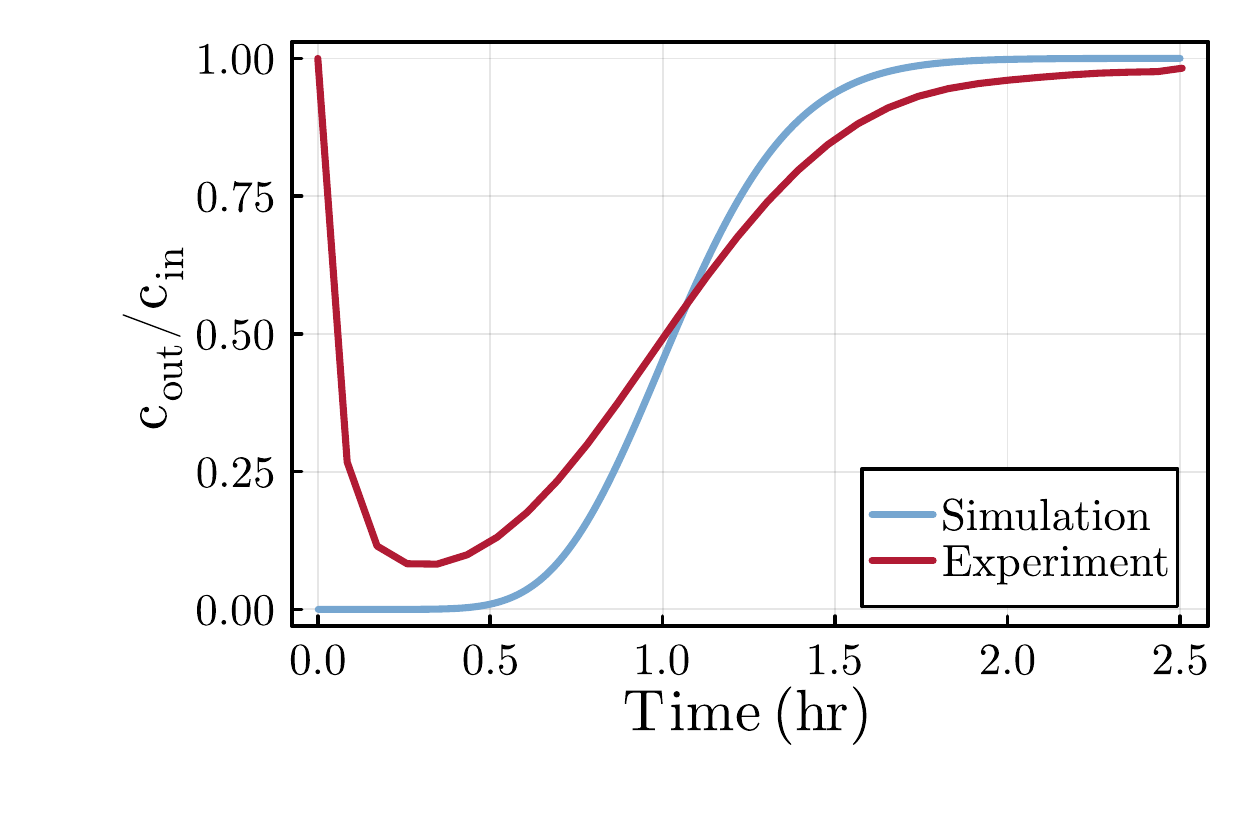}
      \caption{Packed bed of pellets}
      \label{fig:btpb}
  \end{subfigure}%
  \begin{subfigure}{0.5\textwidth}
  \centering
  \includegraphics[width=1\linewidth]{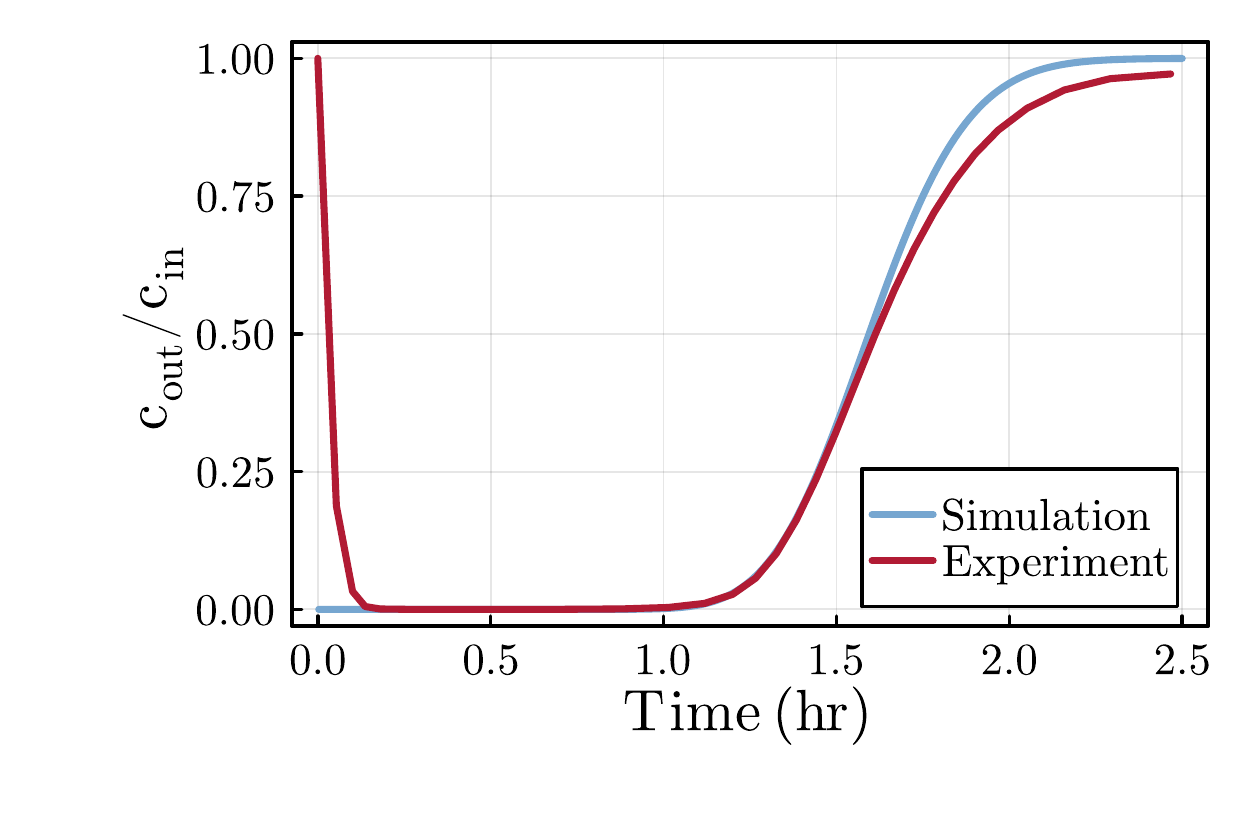}
      \caption{Monolithic contactor}
      \label{fig:btmon}
  \end{subfigure}
  \caption{Validation breakthrough curves}
  \label{fig:bts}
\end{figure}

Breakthrough data is obtained experimentally by following the method in Section \ref{btm} and a smoothed curve is plotted in Figure \ref{fig:bts}. The numerical method in Section \ref{numerical-method} is used considering the same geometrical and physical constants as the experiment, which are listed in Tables \ref{tab:physical-constants} and \ref{tab:geometrical-constants}. The numerical constants used for the breakthrough experiments are seen in Table \ref{tab:numerical-constants}. 
 
The experimental breakthrough curves contain an initial state at which the outlet concentration equals the inlet concentration. This is a feature of the experimental method, where CO\textsubscript{2} readings from the bypass line are initially included in the measurement and should therefore be ignored. Another feature of the experimental setup is the initial zone of the packed bed breakthrough curve which is above zero for the first 20 minutes before a stable inlet concentration is achieved. Considering the remainder of the data to be valid, we obtain a reasonable agreement between the numerical and experimental results. 

\subsubsection{Pressure Drop}

\begin{figure}[H]
  \centering
  \includegraphics[width=0.5\linewidth]{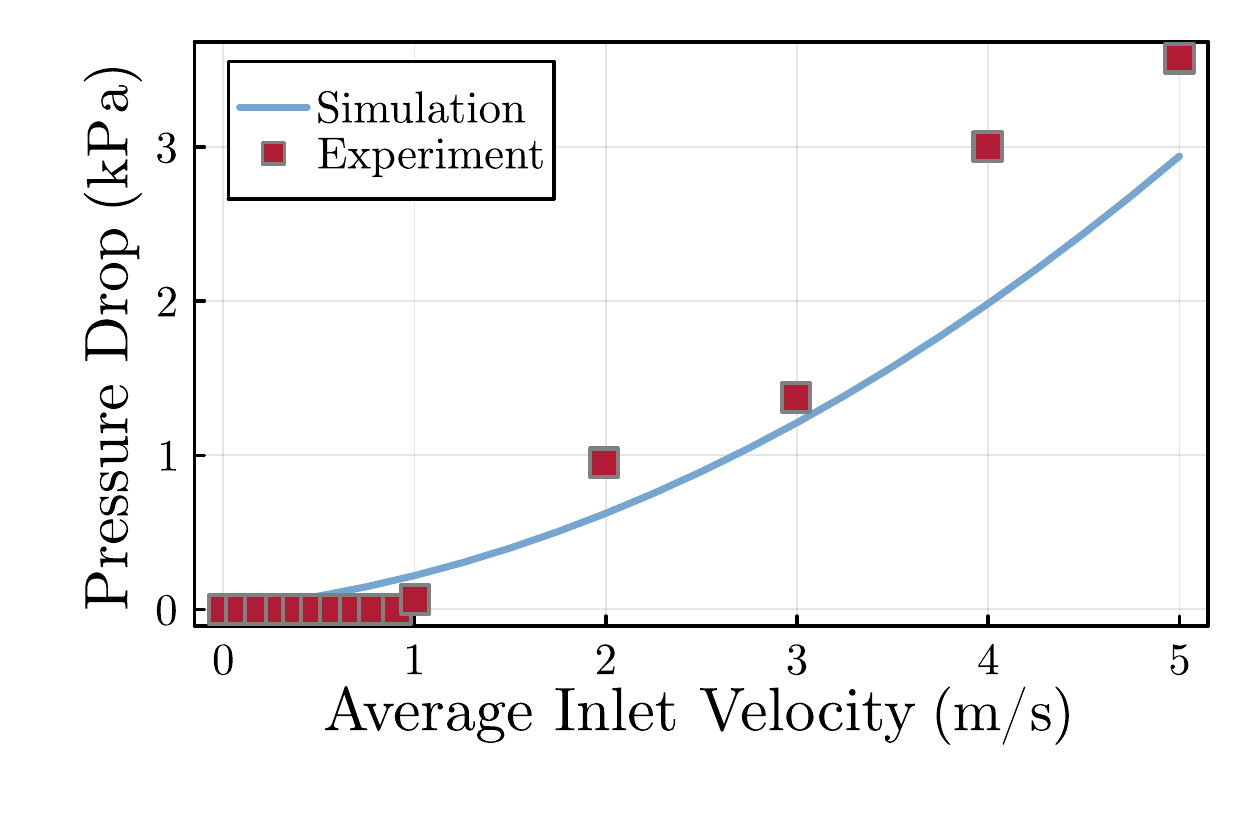} 
  \caption{Validation pressure drops for a varying inlet velocity}
  \label{fig:pd_verification}
\end{figure}

Another critical quantity for a swing adsorption system is the pressure drop across the substrate. This quantity has a large impact on the energy consumption of the system. As seen in Figure \ref{fig:pd_verification}, the pressure drop across the packed bed for a wide range of velocities can be estimated by the simulation.

\hypertarget{performance-evaluation}{%
\section{Performance Derivation}} 
\label{performance-evaluation}

\subsection{Costing Model}

In this section, we describe the costing model proposed by \cite{Sinha2017} which used as a basis for this work. The objective function is taken to be the cost per tonne of CO\textsubscript{2} removed by the system and is made up of various operating and capital components. The total cost in USD per tonne of CO\textsubscript{2} removed by the system is calculated as:
\begin{equation}\label{eq:J}
	J = \sum_{i=1}^6 C_i, 
\end{equation}
where
$C_i$ is the cost (\$/t-CO\textsubscript{2}) associated with component $i$. In this section we outline the different costing components as described in \cite{Sinha2017}.
\nolinebreak
The blower operating cost, $C_1$, is given by:
\begin{equation}
	C_1  = \frac{ Q C_E t_1 dP }{ N_t },
\end{equation}
where
$dP$ is the pressure drop (Pa),
$Q$ is the volumetric flow rate during adsorption (m\textsuperscript{3}s\textsuperscript{-1}),
$C_E$ is the cost of energy (\$J\textsuperscript{-1}),
$t_1$ is the adsorption step time (s) and   
$N_t$ is the amount of CO\textsubscript{2} (tonne) removed per cycle. 
\nolinebreak
{We assume that the sorbent is regenerated by a hot steam as in \cite{Sinha2017}. We note that other regeneration methods, such as resistive heating, are also possible.}
The steam operating cost for the desorption step, $C_2$, is calculated as:
\begin{equation}
	C_2 = \frac{Q_4 t_4 \rho_{st} C_{st} }{ N_t },
\end{equation}
where
$Q_4$ is the volumetric flow rate during desorption (m s\textsuperscript{-1}),
$t_4$ is the desorption step time (s),
$\rho_{st}$ is the density of steam (kg m\textsuperscript{-3}) and 
$C_{st}$ is the cost of steam (\$ kg\textsuperscript{-1}). 
\nolinebreak
The vacuum pump operating cost, $C_3$, assuming an isentropic system is taken as
\begin{equation}
	C_3 = - \frac{C_E}{N_t} \frac{ P_1 V_1^\gamma ( V_2^{1-\gamma} -  V_1^{1-\gamma } ) }{ 1-\gamma },
\end{equation}
where 
$P_1$ is the initial pressure,
$V_1$ and $V_2$ are the initial and final volume during
the vacuum swing step and
$\gamma$ is the adiabatic index.
\nolinebreak
The monolith support capital cost, $C_4$, is taken as:
\begin{equation}
	C_4 = \frac { V_{mon} C_{mon} t_{cycle} } { N_{mon} N_t },
\end{equation}
where
$V_{mon}$ is the monolith volume (\si{m^3}), 
$C_{mon}$ is the cost per volume of monolith (\si{\$ m^{-3}}),
$t_{cycle}$ is the total cycle time (\si{s}) and
$N_{mon}$ is the lifetime of the monolith (\si{s}).
The adsorbent capital cost, $C_5$, is given by:
\begin{equation}
	C_5 = \frac { m_{ads} C_{ads} t_{cycle} } { N_{ads} N_t},
\end{equation}
where
$m_{ads}$ is the adsorbent mass (\si{kg}),
$C_{ads}$ is the cost per mass of monolith (\si{\$/kg}),
$t_{cycle}$ is the cycle time (\si{s}) and 
$N_{ads}$ is the lifetime of the monolith (\si{s}). 
\nolinebreak
The blower capital cost, $C_6$, is given by:
\begin{equation}
	C_6 = \frac { C_B t_{cycle} }  { N_{B} N_t },
\end{equation}
where
$C_{B}$ is the cost of the blower (\$) which is a function of the outlet pressure $p_o$,
$N_{B}$ is the lifetime of the blower (s) and
$t_{cycle}$ is the cycle time (s).
The vacuum pump capital cost, $C_7$, is given by:
\nolinebreak
\begin{equation}
		C_7 = \frac { ( C_{vac} + C_{M} ) t_{cycle} }  { N_{yrB}  N_t }, 
\end{equation}
where
$C_{vac}$ is the cost of the bare module vacuum pump (\si{\$}),
$C_{M}$ is the cost of the electric motor of the pump (\si{\$}) and
$N_{yrvac}$ is the lifetime of the pump (s).

\subsection{Cost Evaluation}\label{cost-evaluation}

To simplify the techno-economic evaluation process and allow for a meaningful comparison to the benchmark presented in \cite{Sinha2017}, we consider, except for the geometry, exactly the same parameters for the \ac{dac} module as in \cite{Sinha2017} using the MIL-101(Cr)-PEI-800 adsorbent. Out of the two adsorbents studied in \cite{Sinha2017}, the MIL-101(Cr)-PEI-800 has a lower uncertainty around the adsorbent capital cost and is therefore the one selected in this work.  In this section, we construct the cost function $J$. To do so, we will look at each $C_i$ individually. 

Firstly, however, we notice that $N_t$ appears in the denominator in every term, so we define this first:
\begin{equation}
	N_t = \alpha_2 M V_f L A, 
\end{equation}
where $\alpha_2=0.7$ is taken from \cite{Sinha2017}, $M$ is the molecular mass in tonne per mol of CO\textsubscript{2},
L is the length of the column,
A is the cross-sectional area of the column and 
$V_f$ is the volume fraction of the bed composed of the adsorbent.

Now we consider the first cost component, $C_1$, the blower operating cost. We notice that it does not matter, for this component, how many channels form the system. In \cite{Sinha2017}, many channels are considered but in ours we have only one. This is irrelevant  because we work out the cost per channel and then divide by the amount of CO\textsubscript{2} per channel. 

Secondly, the first key component that is different to \cite{Sinha2017} is the cross-sectional area of the channel. However, the area of the channels can actually be eliminated from this costing component as we can write $C_1$ as a term multiplied by $A$ divided by a term that is also multiplied by $A$:
\begin{equation}
  C_1 = \frac{ (u_{\text{avg}} C_E t_1 dP) A }{ (\alpha_2 M V_f L ) A }
\end{equation}
The variables that we consider different to \cite{Sinha2017} that impact the cost are the volume fraction $V_f$ of adsorbent within the channel 
and the arrangement of the adsorbent within the channel, which effects the pressure drop, $dP$, and adsorption time, $t_1$.

By considering that only these quantities will change, we can eliminate all the constants in the costing model and only consider variations to the final costing component values as seen in \cite{Sinha2017}. We can do this by rearranging the equations so that each costing component is formulated as the costs seen in \cite{Sinha2017} multiplied by the ratios of the quantities for optimized designs compared to the monolithic baseline design of \cite{Sinha2017}:

Consider the first costing component for the monolith benchmark, $C_{1,s}$ to be computed as:
\begin{equation}
  C_{1,s} = \left(\frac{u_{\text{avg}} C_E}{\alpha_2 M L}\right)\left( \frac{t_{1,s} dP_s}{V_{f,s}}\right).
  \label{eq:c1nt}
\end{equation}
For the monolith in \cite{Sinha2017}, this cost is \$25/t-CO\textsubscript{2}. So, considering the first fraction in Equation \ref{eq:c1nt} to be unchanged, we can eliminate it and instead write out the optimized fraction as:
\begin{equation}
  C_1 = 25 (\frac{t_1}{t_{1,s}})(\frac{dP}{dP_s})(\frac{V_{f,s}}{V_f})
\end{equation}
Now we observe the second costing component, $C_2$. Like $C_1$, we can write $C_2$ in such a way as to cancel out the cross-sectional area and be evaluated as a function of ratios of key parameters only:
\begin{equation}
  C_2 = 15 (\frac{t_4}{t_{4,s}})(\frac{V_{f,s}}{V_f}).
\end{equation}
In this work, we make an approximation that the desorption time scales with the adsorption time by a constant which is geometry independent, so that $t_4/t_{4,s} = t_1/t_{1,s}$.

The third cost component is related to the vacuum pump operation. This cost is very low compared to the entire system \cite{Sinha2017}, so we consider variations negligible and take the value as in \cite{Sinha2017} scaled by the volume fraction only:
\begin{equation}
  C_3 = 2 (\frac{V_{f,s}}{V_f}),
\end{equation}
which is a conservative approximation since we have a lower $P_1$ compared to the monolithic design. 
The fourth cost $C_4$ is the capital cost of the monolithic support. Since we do not have any support in our design, this cost is zero. {The cost of any binder and filler material is considered negligible.}

Next we observe the adsorbent capital cost $C_5$. The cost of the adsorbent material contains significant uncertainties, so, as in \cite{Sinha2017}, we take a range of possible values for the quantity $C_{ads}$. We can express the upper and lower bound estimates $C_{5,L}$ and $C_{5,L}$ as a function of the key parameters as:
\begin{equation}
\begin{aligned}
	C_{5,L} &= 5 (\frac{t_1}{t_{1,s}}) \quad  \text{and} \\
	C_{5,U} &= 70 (\frac{t_1}{t_{1,s}}).
\end{aligned}
\end{equation}
Note here there is no dependence on the volume fraction since it appears in both the numerator and denominator.
The purchase cost of the blower is taken as in \cite{Sinha2017}. This is a conservative approximation because of the lower pressure drop across the channel in the optimized design compared to \cite{Sinha2017}. The capital cost $C_6$ is then given by:
\begin{equation}
  C_6 = 15 (\frac{t_1}{t_{1,s}})(\frac{V_{f,s}}{V_f}) ,
\end{equation}
Similarly, the purchase cost of the vacuum is conservatively taken as in \cite{Sinha2017} and the cost $C_7$ can be evaluated as:
\begin{equation}
  C_7 = 9 (\frac{t_1}{t_{1,s}})(\frac{V_{f,s}}{V_f}).
\end{equation}

\subsection{Key Quantities}
{As seen in Section \ref{cost-evaluation}, the cost $J$ is a function of three quantities which vary based on the geometrical layout of the adsorbent. These are
the volume fraction $V_f$ of adsorbent within the channel, the pressure drop $dP$, and adsorption time $t_1$. 
We will now connect the \ac{fe} problem to these key quantities.
To evaluate the performance of the system, we first solve the flow and transport problems defined by \eqref{eq:rins} and \eqref{eq:rad} to obtain $\boldsymbol{u}_h,p_h,c_h$ and $q_h$.
Then, the key quantities for measuring performance can be computed.} 
Firstly, based on the geometry alone we can compute $V_f$:
\begin{equation}
  V_f = (1-\epsilon_{\text{macro}}) \frac{\int_{\Omega} \psi \  d\Omega.}{\int_{\Omega} \  d\Omega}.
\end{equation}
The pressure drop is computed as:
\begin{equation}
  dP = \int_{\Gamma_{in}}p_h \ d\Gamma - \int_{\Gamma_{out}}p_h \ d\Gamma,
  \end{equation}
and the adsorption time is calculated as the time it takes for the adsorbent to reach $\%90$ capacity where the capacity is calculated considering full availability of CO\textsubscript{2} everywhere in the solid, that is, $q=q*$. We then are required to find $t_1$ such that:
\begin{equation}
  \int_{\Omega_\text{s}} q_h\vert_{t=t_1} \ d\Omega = 0.9\int_{\Omega_\text{s}} q^*(p,\bar{c}) \ d\Omega.
\end{equation}
Now, we have completely defined a means to compute the cost of a given geometrical layout.

\section{Optimization Problem}\label{optimization-problem}

We aim to solve the following optimization problem:
\begin{equation}
  \begin{aligned}
    \underset{\mathbf{p}}{\text{min}} & \ J(\boldsymbol{\phi}_h(\mathbf{p}),\mathbf{p})&  \\
    \text{s.t} & \ \mathscr{R}(\boldsymbol{\phi}_h(\mathbf{p}),\mathbf{p}) &= 0 ,\\
         & \ \quad \quad \  \mathscr{V}(\mathbf{p})     &= 0 ,\\
  \end{aligned}
\label{eq:opt-problem}
\end{equation}
where 
  $\mathbf{p}$ are the parameters that describe our geometry,
  $\mathscr{V}$ is an inequality constraint for the volume,
  $\mathscr{R}$ is the total \ac{pde} residual {given by the sum of the expressions in \eqref{eq:rins} and \eqref{eq:rad}},
  $\boldsymbol{\phi}_h = (\boldsymbol{u}_h,p_h,c_h,q_h)$ is the weak solution of the \ac{pde} residual $\mathscr{R}$ and
  $J$ is the objective {defined in \eqref{eq:J}}. Throughout the remainder of this section, we detail the various components of the optimization problem.


\subsection{Modified Topology Optimization Problem}

We could perform the optimization considering the time-dependent problem, although many time-steps are required. This results in a large computational burden on the optimization process due the large computational cost of computing time-dependent adjoints and, in this particular case, is not necessarily critical for finding performant optimized topologies. Instead, we can use a single solution at a fixed time $\bar{t}$ to find update directions for the geometry. We do this by taking a single time-step from the initial condition and solve the pseudo-steady problem by taking:
\begin{equation}
\begin{aligned}
\frac{\partial c}{\partial t} = \frac{c-c\mid_{t=0}}{\bar{t}}\ \ \ \   \text{and} \ \ \ \frac{\partial q}{\partial t } &= \frac{q-q\mid_{t=0}}{\bar{t}}.
\end{aligned}
\end{equation}

We also only use the non-convective residual, i.e. the Brinkman-Forcheimer problem residual, to aid the convergence of the optimizer. Although this approach only provides an approximation of the solution, we show that, by testing the optimized geometries using the complete convective Brinkman-Forcheimer problem and transient transport formulation, the geometries obtained using the auxiliary problem provide improved performance over the standard designs. Note that $\bar{t}$ is a numerical parameter and, because we are not capturing the dynamics of the full problem, it has no physical meaning.

For the one-step optimization approach, we define a modified objective $j$ and allow the value of $\bar{t}$ to be a hyperparameter in the optimization. After obtaining optimized geometries, we compute the final performance measures based on the full transient simulation using the original objective $J$. The modified topology optimization problem can then be stated as: 
\begin{equation}
  \begin{aligned}
    \underset{\mathbf{p}}{\text{min}} & \ j(\boldsymbol{u}(\mathbf{p}),\mathbf{p})&  \\
    \text{s.t} & \ \mathscr{R}_s\mid_{\bar{t}}   (\boldsymbol{u}(\mathbf{p}),\mathbf{p}) &= 0 ,\\
    & \ \quad \quad \  \mathscr{V}(\mathbf{p})     &< 0 ,\\
  \end{aligned}
\label{eq:opt-problem}
\end{equation}
where $\mathscr{R}_s$ is the pseudo-steady non-convective form of $\mathscr{R}$ and
 $j$ is the sum of weighted objectives:
  \begin{equation}
  j = j_T + w_1 j_E
\end{equation}
where $w_1$ is a scaling for the two terms. 
The modified objective is designed to impact the key quantities $dP$, $dq$ and $t_1$. The pressure drop is taken as the component $j_E=dP$ and the second component, $j_T$ is computed as the amount adsorbed at $\bar{t}$:
\begin{equation}
  j_T =  -\int_{\Omega} q\mid_{\bar{t}} d\Omega,
\end{equation}
where the minus sign is included because this is a quantity that we wish to maximize. {Maximizing this quantity has an impact on the quantity $\bar{t}$ as if there is a greater adsorbed amount at a given time, it is likely that the adsorbent will be faster to reach capacity.}

\subsection{Geometry Parameterization}
Following the \ac{SIMP} method for topology optimization \cite{Sigmund2001} and enforcing admissibility of optimized designs, we define the map between the design variables $\mathbf{p}$ and the interpolation function $\psi$.

The design variables represent element-wise properties of the design area and take on a value of $1$ if the element contains the adsorbent and $0$ if the element is void. We therefore use the piece-wise constant space $V^0_h$, which is isomorphic with $\mathbb{R}^N$, to define the field $\psi_p \in V^0_h$ by the degrees of freedom $\mathbf{p}\in\mathbb{R}^N$.

The first processing step is to use a Helmholtz filter \cite{Lazarov2010} to smooth and prevent oscillations in the geometry which is done by finding the solution $\psi_f$ to the Equation:
\begin{equation}
  r_f^2 \nabla \psi_f \nabla v_p + \psi_f v_p = \psi_p v_p 
\end{equation}
for all $v_p \in V^1_h$ where $r_f$ gives the smoothing radius. 

\subsection{Penalization}
{Standard in density-based topology optimizations, the penalization of intermediate densities is used to ensure designs that are comprised of free flow, i.e. $\psi=0$, regions and porous flow, i.e. $\psi=0$, regions with only a small region of indicator values between 0 and 1 in the vicinity of the interface. In this work, we take 
\begin{equation}
  \psi=\psi_f^5
\end{equation} 
to promote these types of 0-1 solutions}
Also, we find that 0-1 designs are promoted by multiplying the isotherm by the indicator field, i.e. $ q^* = \psi K c $. Note that this is inconsistent in the grey area, so we remove this factor when considering the full transient simulation and only include it in the optimization. 

\subsection{Constraints}
  With the parameter to objective map defined, we only are left to specify the volume constraint for the problem:
  \begin{equation}
    \mathscr{V} = \frac{\int_{\Omega} \psi_f \mathrm{d}\Omega}{  \int_{\Omega} \mathrm{d}\Omega} - V_{max} < 0,
  \end{equation}
  where $V_{max}$ is the maximum allowed volume fraction for the porous region in the bed.


\section{Optimized designs}\label{optimized-designs}

In this section, we perform the \ac{to} of the bed. 
{We utilize Gridap to solve the \ac{fe} problem and evaluate the objective function but now also make use of the 
Method of Moving Asymptotes \cite{Svanberg1987} algorithm with 400 iterations to perform the optimization.}
Note that here we consider an optimization of the macrostructure of the bed. That is, we investigate the divide between a free-flow region and a porous region. We do not optimize the microstructure of the porous material. What we do, instead, is consider the porous structure to be made up of spherical beads. We choose spherical beads because the homogenization of this microstructure to obtain macroscopic properties is well studied and considered robust \cite{Erdim2015,Nemec2005,Shafeeyan2014}. The Reynolds number for the flow through the porous region in the optimized design is around 200 and in the range of what is considered valid for the Ergun equation \cite{Erdim2015}. The packed sphere arrangement is certainly not considered, however, to be an optimal microstructure. It has been shown, in fact, that alternative microstructures can offer much lower pressure drops \cite{Dhoke2021}. Alternative microstructure would therefore likely lead to better performance. 

Considering spherical beads for the design, we are free to select the bead size and porosity. We select the radius of the beads equal to the pellets verified in Section \ref{validation}. Using beads of a given radius implies a restriction on the feature size of the geometry. We should only consider features that are of a greater length scale than the radius of the beads to ensure admissibility of the optimized designs. To do this, we use a filter radius, $R_f$, of 0.5 during the optimization steps and a filter radius of 6 in a post-processing step. We find that this combination prevents fine scale features on the order of the bead radius. The final geometry is also passed through a threshold for a sharper interface between the free flow and porous regions. To select the porosity, we assume a regular arrangement of spheres, which corresponds to a porosity of 0.476 \cite{vonSeckendorff2021}. This defines the maximum porosity for packed spheres. Beyond this porosity, the spheres would be disconnected. 

In order to verify the effectiveness of the approach, we consider a comparison to baseline solutions. These include the monolithic arrangement see in \cite{Sinha2017} and a packed bed of spheres. We consider the model in \cite{Sinha2017} to be reasonable for the simple monolithic arrangement and take their results as the benchmark. We also consider the case of a completely full packed bed with a porosity and bead size equal to that taken for the microstructure in the optimized designs. The completely full packed bed corresponds to taking an interpolation function, $\psi$, equal to 1 everywhere in the design domain.

\begin{table}[htbp]
  \centering
  \caption{Optimization parameters}
  \begin{tabular}{lll}
  \hline
  Constant & Value & Unit \\
  \hline
  $w_1$ & 70 & - \\
  $\bar{t}$ & 80 & s \\ 
  $\epsilon_\text{bed}$ & 0.476 & - \\ 
  $d_p$ & 3.2 & mm \\ 
  $V_{\mathrm{max}}$ & 0.95 & - \\
  \hline
  \end{tabular}
  \label{tab:geometrical-constants}
\end{table}

\begin{figure}
	\centering
  \includegraphics[width=1\linewidth]{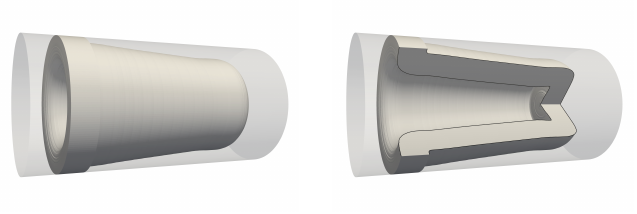}
	\caption{Optimized geometry: On the left is the full rendered design and on the right is a segment of the design}
	\label{fig:optgeo}
\end{figure}

\begin{figure}
	\centering
  \includegraphics[width=1\linewidth]{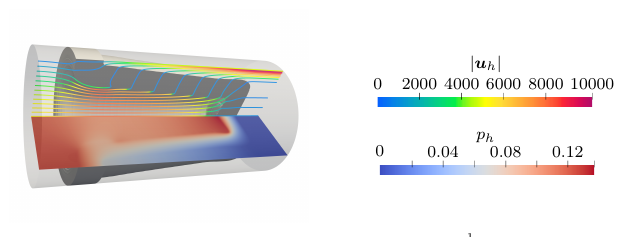}
	\caption{Fluid solution: The solution of the convective Brinkman-Forcheimer equations for the optimized design.}
	\label{fig:optu}
\end{figure}

The optimized design in Figure \ref{fig:optgeo} resembles a hollow blunted cone which involves short and even fluid pathways through the porous material, as seen by the streamlines in Figure \ref{fig:optu}. The pressure drop in the void region is very small compared to the solid. The majority of the pressure drop occurs across the solid region, with only small amounts of dissipation in the void. Having a lower fluid velocity and a short fluid pathway across the solid domain is thus what allows for a low pressure drop for this geometry. 

\begin{figure}
	\centering
  \includegraphics[width=1\linewidth]{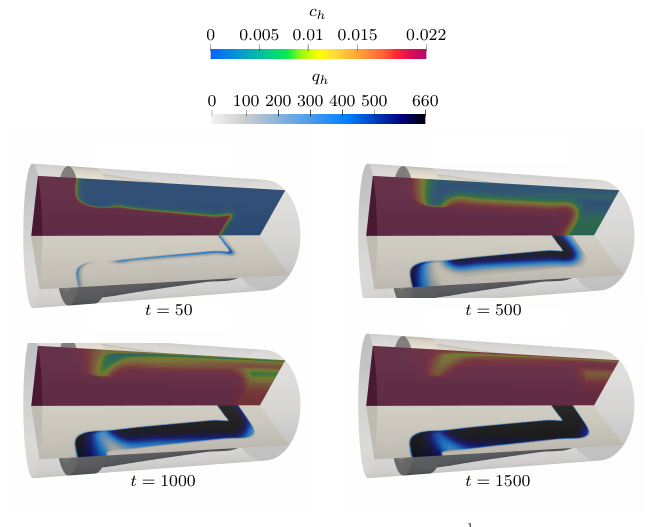}
	\caption{Transport solution: The solution of the transport equations with adsorption in the solid domain.}
	\label{fig:optc}
\end{figure}

\begin{figure}
	\centering
  \includegraphics[width=0.5\linewidth]{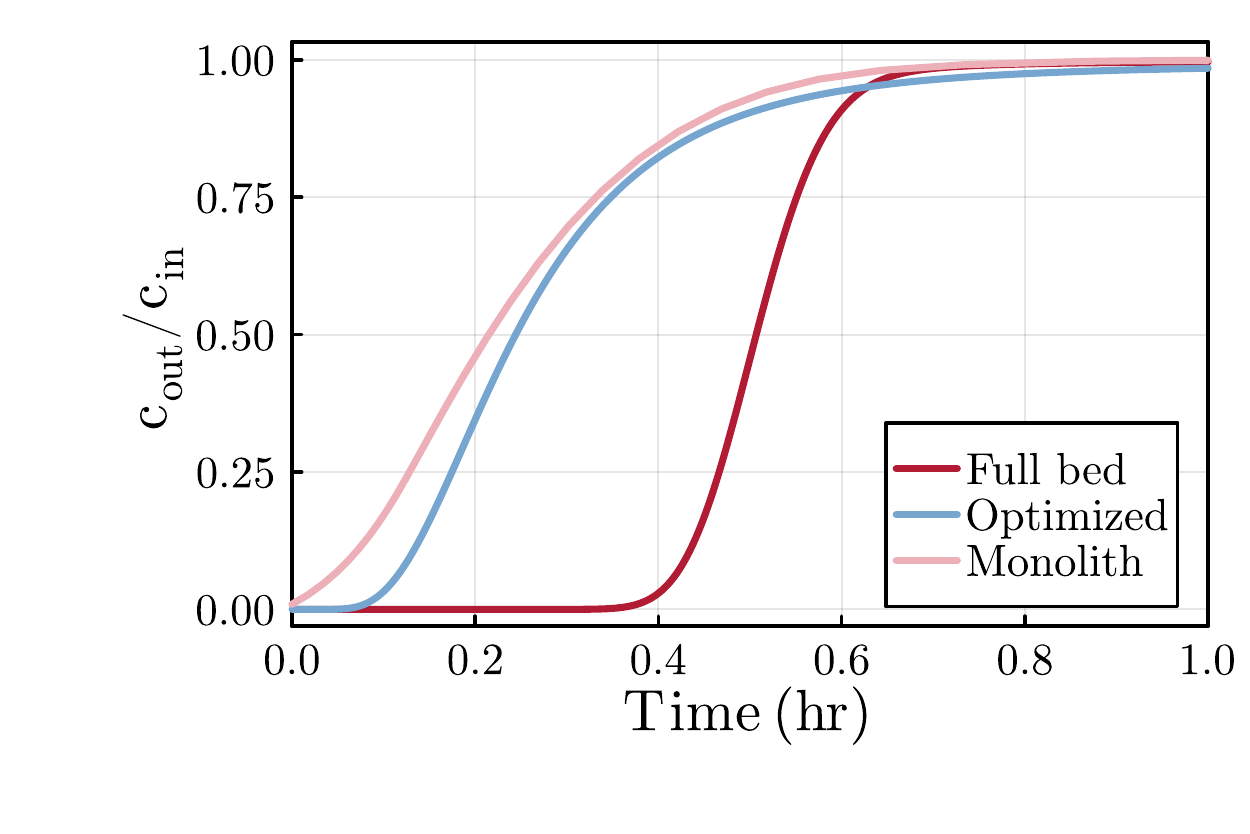}
	\caption{{Breakthrough curves for the alternate designs.}}
	\label{fig:bt_comp}
\end{figure}

The arrangement also ensures that the CO\textsubscript{2} gets distributed through the solid so that the concentration front does not spread out over a large time. We can see in Figure \ref{fig:optc} that the CO\textsubscript{2} front moves across the solid so that the concentration downstream is always within a small range. This is mainly because the transport pathways through the solid structure are generally of even lengths. {Consequently, the breakthrough curve has a similar shape to the monolithic arrangement, as can be seen in Figure \ref{fig:bt_comp}, although we see significant performance improvements because of the lower pressure drop.}

By using the optimized arrangement, the pressure drop is significantly reduced and the breakthrough time remains small despite having more adsorptive material compared to the monolithic arrangement. This allows for a cost reduction from the \$75-140/t-CO\textsubscript{2} reported in \cite{Sinha2017} to a range of \$49-116/t-CO\textsubscript{2}. Note that the range is due to the uncertainty around the capital cost of the adsorption material. The full packed bed is associated with a significantly larger pressure drop and therefore a much larger cost of \$267-357/t-CO\textsubscript{2}.  

\begin{table}[htbp]
  \centering
  \caption{Objective functions for different geometrical layouts}
  \begin{tabular}{l c c c c}
  \hline
  \multicolumn{1}{p{5cm}}{\centering Design } & \multicolumn{1}{m{3cm}}{\centering $dP$(Pa)} & \multicolumn{1}{m{1cm}}{$t_1$(s)} & $V_{\text{ads}}$ & {$J$(\$/t-CO\textsubscript{2})} \\
  \hline
  monolith                  & 433 & 1150 & 0.21 & 75-140 \\
  full packed bed           & 7329 &  1600 & 0.52 & 267-357 \\
  \bf{optimized packed bed} & 124 & 1180 & 0.24 & 49-116 \\
  \hline
  \end{tabular}
  \label{tab:jj_m}
\end{table}

\section{Conclusion}

A numerical model for the simulation of a \ac{dac} system was proposed. To validate the complete numerical model, physical breakthrough experiments and pressure drop tests using different geometrical layouts of the adsorbent bed were conducted. A reasonable fit between the experimental and simulated data was found. The validated numerical model was then embedded within a \ac{to} framework for the purpose of discovering new geometrical layouts of the adsorption bed.
The optimization of the macrostructure of the bed unveiled a simple blunted cone design associated with a low pressure drop and fast mass transfer kinetics. The proposed structure offers improved performance compared to benchmark solutions and could possibly be useful more generally for adsorption and catalysis applications where low pressure drops and high contact for a chemical species is desired. 

In this work, the microstructure was assumed to be composed of packed spheres because of the robust homogenization models available. However, a topology optimization of the microstructure itself would also likely lead to further improvement to the performance of the system.


\bibliographystyle{unsrtnat} 
\bibliography{/home/mallon2/Documents/ConnorMallon-Thesis-1/doc/thesis/Preamble/bibliography.bib}

\end{document}